\documentclass[aps,pra,twocolumn,showpacs,superscriptaddress,groupedaddress]{revtex4}

\usepackage{amssymb}
\usepackage{color,graphicx}
\usepackage{amsmath}
\usepackage{amsbsy}
\usepackage{amsthm}
\usepackage{bbm}
\usepackage{bm}
\usepackage{epsfig}
\usepackage{lscape}

\newcommand{\ket}[1]{\left| {#1} \right\rangle}
\newcommand{\bra}[1]{\left\langle {#1}\right|}

\newcommand{\abs}[1]{\left| {#1} \right|}
\newcommand{\ve}[1]{\mathbf{#1}}

\begin{document}

\title{Coherent scattering in non relativistic quantum mechanics}

\author{Giulio Gasbarri}
\email{giulio.gasbarri@ts.infn.it}
\affiliation{Department of Physics, University of Trieste, Strada Costiera 11,
34151 Trieste, Italy}
\affiliation{Istituto Nazionale di Fisica Nucleare, Trieste Section, Via Valerio
2, 34127 Trieste, Italy}

\author{Sandro Donadi}
\email{sandro.donadi@ts.infn.it}
\affiliation{Department of Physics, University of Trieste, Strada Costiera 11,
34151 Trieste, Italy}
\affiliation{Istituto Nazionale di Fisica Nucleare, Trieste Section, Via Valerio
2, 34127 Trieste, Italy}

\author{Angelo Bassi}
\email{bassi@ts.infn.it}
\affiliation{Department of Physics, University of Trieste, Strada Costiera 11,
34151 Trieste, Italy}
\affiliation{Istituto Nazionale di Fisica Nucleare, Trieste Section, Via Valerio
2, 34127 Trieste, Italy}

\date{\today}

\begin{abstract}
We show under which conditions a particle scatters coherently on a multi-particles system, working in the non relativistic framework. In a nutshell, in order to have coherent scattering, the incident particle has to not resolve the internal structure of the composite system. We show that the above condition is satisfied when the de Broglie length of the incident particle is much larger than the size of the system.
\end{abstract}

\maketitle

\section{Introduction}
Aim of this paper is to study coherent scattering in the non relativistic regime.
In general, when a particle is scattered by a composite system (e.g. an electron hitting an atom), the scattering depends in a non-trivial way on the interaction of the incident particle with each of the particles of the target. However,  under certain conditions, all particles of the target scatter the incident particle in the same way, leading to ``coherent scattering". 

As we will show, coherent scattering arises when the incident particle does not resolve the internal structure of the target. In such a case, in first approximation, the target can be treated as a point like object with total charge equal to the sum of the charges of its constituents \footnote{We wish to clarify that, since the results we will obtain are general, here we refer to generic charges, not necessarily the electric charges}. Coherent scattering is interesting, for the following two properties: ({\it i}) it does not require a detailed knowledge of the internal structure of the target and ({\it ii}) when all the $N$ constituents have the same charge, the cross section scales as $N^2$ and, therefore, is larger compared to the case of incoherent scattering, where the cross section increase as $N^{\alpha}$ with $1\leq\alpha<2$ ($\alpha=1$ corresponds to the total incoherent scattering, where all the components of the target scatter independently, while for $1<\alpha<2$ the scattering is partially coherent and partially incoherent).

Property ({\it i}) played a fundamental role on a very important and famous experiment of last century performed by Ernest Rutherford and his collaborators. They fired energetic $\alpha$-particles to foils of different materials, and measured the deflection of the scattered particles~\cite{Geiger,rutherford2012scattering}. From the distribution of scattered particles they deduced information on the structure of the foil and were able to prove the planetary model of the atom.
The model used to fit the experimental data is very simple: the $\alpha$-particle and the nucleus are described as point-like particles and the details about their internal structure are ignored, exactly as in the coherent scattering regime. As a result, the cross-section scales with the square of the number of protons in the nucleus. This square dependence is a clear sign that all the protons in the nucleus scatter coherently off the incident $\alpha$-particle.

Property ({\it ii}) motivates our study on coherent scattering in the non relativistic regime. In fact, different interferometry experiments in this regime have been recently proposed, where the interactions considered are usually very weak~\cite{bateman2014existence, Riedel}. Therefore, a quadratic increase of the cross section due to the presence of coherence may play an important role in making these experiments more effective.
  
The main result of this paper is to show under which conditions coherent scattering arises, i.e. when the De Broglie wave length of the incident particle is much larger than the typical size of the target.


The paper is organized as follows: in section~\ref{sec-exact} we study a simple model where a particle moving in one dimension is scattered by two Dirac delta potentials and we derive the conditions under which there is coherent scattering. In section~\ref{sec-scattering} we repeat this analysis by studying, in the perturbative regime, a more realist situation of scattering between a particle and a generic N-particle bound system, through a generic potential. In section~\ref{sec-cross} we derive the formula for the cross section in the coherent scattering regime. 
In section~\ref{sec-ruth} we apply the coherent scattering cross section formula to the case of the scattering of $\alpha$-particle off nucleus of different atoms, re-obtaining the Rutherford cross section.
To conclude, in section~\ref{sec-conc} we summarize the main results of the paper. 


\section{Coherent scattering: an exact calculation}\label{sec-exact}

We start our analysis by studying a simple model where all calculations can be carried out exactly. We consider the one dimensional scattering of a particle off two Dirac delta potentials. 
We start with this simple example in order to avoid unnecessary mathematical difficulties. We will get a clear picture of the conditions under which coherent scattering occurs.
The dynamics of the system is described  by the Schr{\"o}dinger equation with Hamiltonian:
\begin{equation}
\label{eq:HHH}
{\hat H}=\frac{{\hat p}^{2}}{2m}+\alpha\,\delta\left({\hat x}\right)+\alpha\,\delta\left({\hat x}-a\right)
\end{equation}
where $m$ is the mass of the particle and $\alpha$ is the coupling constant of the $\delta$-potentials (with dimension
$\left[\alpha\right]=$J m). 
As it is shown in Appendix A, the scattering probability $R$ for this simple model is given by:
\begin{equation}
R=\frac{4\left[\beta\cos(ka)+\sin(ka)\right]^{2}}{2+2\beta^{2}+\beta^{4}+2(\beta^{2}-1)\cos(2ka)+4\beta\sin(2ka)}
\label{riflesso}
\end{equation}
where $\beta=\frac{k\hbar^{2}}{m\alpha}$.
The scattering probability $R$ depends on the wave vector $k$ of the incident plane wave and on the distance $a$ between the two delta potentials.
We expect to observe coherent scattering when the particle's wave length $2\pi/k$ is much larger than $a$ (which means $ka\ll1$). In fact, in this regime, the distance between the two centers is small enough that the two scattered components of the wave come out with almost exactly the same phase, adding therefore coherently.
In other words, the plane wave representing the incident particle does not resolve the distance $a$ between the two potentials and therefore it scatters off the two deltas as if they were one on top of the other.
To verify this, let us approximate the reflection probability $R$ of Eq.~\eqref{riflesso} in the regime $ka\ll1$. 
In this regime Eq.~\eqref{riflesso} becomes:
\begin{equation}
R\simeq\frac{4}{4+\beta^{2}}=\frac{1}{1+\frac{k^{2}\hbar^{4}}{m^{2}(2\alpha)^{2}}}.\label{riflesso cohe}
\end{equation}
This result should be compared with the reflection probability formula for a scattering process with only one delta potential in the origin:
\begin{equation}
R_{single}=\frac{1}{1+\beta^{2}}=\frac{1}{1+\frac{k^{2}\hbar^{4}}{m^{2}\alpha^{2}}}.\label{riflesso single}
\end{equation}
We see that Eq.~\eqref{riflesso cohe} describes the same reflectivity as that of a single delta with coupling constant $2\alpha$. In the regime $ka\ll1$ the incident plane wave is not able to distinguish the two $\delta$-potential and their effect adds coherently. \\  

Something interesting happens when the interaction is weak and perturbation theory can be applied.
 In the perturbative regime, the velocity of the particle is much larger than the coupling constant of the interaction i.e. $\frac{\hbar k}{m}\gg \frac{\alpha}{\hbar}$, which is equivalent to having $\beta\gg1$.
Then the denominators in Eqs.~(\ref{riflesso cohe}) and (\ref{riflesso single}) can be approximated
by $\beta^{2}$ and we have:
\begin{equation}
R\simeq 4\, R_{single}.
\end{equation}
The factor $4$ is exactly the contribution due to the presence of the two deltas. As we can see, when the Born approximation applies, the probability of having a scattering (which in this case is given by the reflectivity $R$) increases with the square of the number of the scattering centers. This, as we will show in the next sections, is true not only for this simple model, but also for more generic systems and interactions.
Therefore, when the interaction is weak, the coherent scattering regime can be used to increase quadratically the cross section, and with it the measurable effects.

\section{Modeling the scattering process}\label{sec-scattering}
In this section we extend the previous results to a more generic situation:
The scattering of a particle  off a bound system of $N$ particles (which in the following we will refer to this as ``target"), in the non relativistic regime.
We first define the target, the incident particle and the interaction.
\subsection{The Target}
We consider a bound system of $N$ interacting particles (e.g. an atom or a molecule), which is described by the following Hamiltonian:
\begin{equation}\label{H0}
\hat{H}_{0}=\sum_{i=1}^{N} \frac{\hat{\ve{q}}_{i}^{2}}{2m_{i}}+\sum_{i<j=1}^{N}\alpha_{ij}U(\hat{\ve{x}}_{i}-\hat{\ve{x}}_{j})
\end{equation}
where $\hat{\ve{q}}_{i}$ is the momentum operator of the $i$-th particle, $\hat{\ve{x}}_{i}$ its position operator, and the coupling constants $\alpha_{ij}$ measure the strength of the interaction $U$ between the $i$-th and the $j$-th particle.
It is convenient to introduce the center of mass and relatives coordinates, which are given by:
 \begin{eqnarray}
\begin{array}{lll}
\text{\small Position of Center of Mass  }&\hspace{0.1cm}\displaystyle \hat{\ve{X}}= \sum_{i=1}^{N}\frac{m_{i}}{M}\hat{\ve{x}}_{i}\\
\\
\text{\small Total Momentum } & \hspace{0.2cm}\displaystyle \hat{\ve{P}}= \sum_{i=1}^{N}\hat{\ve{q}_{i}}  \\
\\
\text{\small Relative Positions} & \hspace{0.2cm}
\left\{\begin{array}{lll}
\hat{\bm{r}}_{i}=\hat{\ve{x}_{i}}-\hat{\ve{X}} 
\\
\displaystyle \hat{\bm{r}}_{N}=-\sum_{i=1}^{N-1}\frac{m_{i}}{m_{N}}\hat{\bm{r}}_{i}
\end{array}
\right.
\\
\nonumber\\
\text{\small Relative Momenta} & \hspace{0.2cm}
\left\{
\begin{array}{lll}
\displaystyle \hat{\ve{p}_{i}}=\hat{\ve{q}}_{i}-\frac{m_{i}}{M}\hat{\ve{P}}
\nonumber\\
\displaystyle \hat{\ve{p}}_{N}=-\sum_{i=1}^{N-1}\hat{\ve{p}_{i}} \nonumber
\end{array}
\right.
 \end{array}\\
 \label{eq:ptrr}
\end{eqnarray}
where $i\in(1,\dots N-1)$  and $M=\sum_{i=1}^{N}m_{i}$ the total mass of the system. Using these coordinates, the Hamiltonian $\hat{H}_{0}$ of Eq.~\eqref{H0} can be written as
\begin{align}
\hat{H}_{0}&{}=\frac{\hat{\ve{P}}^{2}}{2M}+\sum_{i=1}^{N} \frac{\hat{\ve{p}}_{i}^{2}}{2m_{i}}+\sum_{i<j=1}^{N} \alpha_{ij}U(\hat{\ve{r}}_{i}-\hat{\ve{r}}_{j}),
\end{align} 
in which the separation between the center of mass degrees of freedom and the internal degrees of freedom is explicit.
 \subsection{Incident particle and Interaction Hamiltonian}
We consider an incident particle with no internal structure, described by the free Hamiltonian $\hat{H}_{d}$, which interacts with the target through interaction Hamiltonian $\hat{H}_{I}$:
\begin{align}\label{chenometido}
\hat{H}_{d} &= \frac{\hat{\ve{p}}^{2}_{d}}{2m_{d}}, 	&\hat{H}_{int} &= \sum_{i=1}^{N}g_{i}V(\hat{\ve{Y}}-\hat{\ve{X}}-\hat{\bm{r}}_{i}).
\end{align}
In the above equation $g_{i}$ the coupling constants of the interaction between the incident particle and the $i$-th particle of the target, $\hat{\ve{Y}}$ the position operator of the incident particle and $\hat{\ve{X}}$ and ${\hat{\bm{r}}}_{i}$ are, respectively, the center of mass of the target and the relative coordinates of the $i$-th particle, introduced above.

\subsection{The Scattering Process}\label{sct}

According to scattering theory~\cite{taylor2012scattering}, the transition probability  $P_{in\rightarrow out}$ from an initial state $\ket{In}$ to a final state $\ket{Out}$ is given by:
\begin{align}\label{eq:Tprobability}
P_{in \rightarrow out} =\abs{\bra{Out}\hat{T}\ket{In}}^{2},
\end{align}
where $\hat{T}$ is the transition matrix defined via the scattering matrix as $\hat{S}=1+\hat{T}$. 

When the interaction is weak, we can stop the expansion at the first perturbative order (Born Approximation) and $\hat{T}$ becomes~\footnote{In order to use the Born approximation, we must be in the perturbative regime. This is true when $\frac{\bar{V}R}{\hbar}\ll \frac{h\abs{\ve{p}}_{r}}{m_{r}} = \frac{\hbar^{2}}{\lambda_{p_{r}}m_{r}}$
 where R is the linear distance in which the potential is appreciably different from zero, $\bar{V}$ is the mean potential in this region and ${\ve{p}}_{r}$ and $m_r$, respectively, the relative momentum and mass between the target and the incident particle~\cite{Stenko}.}:
\begin{align}\label{eq:Born}
\hat{T}{}&\simeq \lim_{T\rightarrow\infty}-\frac{i}{\hbar}\sum_{i=1}^{N}g_{i}\int_{-T}^{+T}d\tau \,V_{I}(\hat{\ve{Y}}-\hat{\ve{X}}-\hat{\bm{r}}_{i};\tau)
\end{align}
Where $\hat{V}_{I}$ is the potential introduced in Eq.~\eqref{chenometido} in the interaction picture.
The initial and final states $\ket{In}$ and $\ket{Out}$ are usually chosen to be eigenstates of the free Hamiltonian: \begin{eqnarray}
\ket{In}&=&\ket{\ve{p}_{d};\ve{P};\phi}\\
\nonumber\\
\ket{Out}&=&\ket{\ve{p'}_{d};\ve{P'};\phi'}
\end{eqnarray}
where $\ve{p}_{d}$ is the momentum of the incident particle, $\ve{P}$ the momentum associated to the center of mass of the target and $\phi$ denotes an eigenstate of the Hamiltonian relative to the internal degrees of freedom of the target. Note that in this section we consider the initial state of the incident particle to be described by a plane wave. However, the extensions to the case of an incident particle described by a wave packet or the case of a beam of incident particles described by a statistical operator lead to very similar results. Appendix B and appendix C contain an explicit calculation.    
The matrix element of the $\hat T$ operator defined in Eq.~\eqref{eq:Born} can be written as follows:
\begin{widetext}
\begin{align}\label{eq:s1new}
&\bra{\ve{p}_{d}',\ve{P}',\phi'}\hat{T}\ket{\ve{p}_{d},\ve{P},\phi} = \nonumber\\
&\qquad \qquad=\lim_{T\rightarrow\infty}-\frac{i}{\hbar}\sum_{i=1}^{N}g_{i} \int _{-T}^{T} d\tau \, e^{\frac{i}{\hbar}(E-E')\tau}\int d\ve{Y} \frac{e^{\frac{i}{\hbar}(\ve{p}_{d}-\ve{p}'_{d})\cdot\ve{Y}}}{(2\pi\hbar)^{3}}\int d\ve{X} \frac{e^{\frac{i}{\hbar}(\ve{P}-\ve{P}')\cdot\ve{X}}}{(2\pi\hbar)^{3}} \int d\{\bm{r}\}V(\ve{Y}-\ve{X}-\bm{r}_{i})\phi'(\{\bm{r}\})^{*}\phi(\{\bm{r}\})\nonumber\\
&\qquad \qquad =\lim_{T\rightarrow\infty}\frac{-i}{(2\pi)^{2}\hbar^{3}} \, \delta_{T}(E-E')\delta(\ve{p}_{d}+\ve{P}-\ve{p}'_{d}-\ve{P}') \tilde{V}(\ve{p}_{d}-\ve{p}'_{d})\sum_{i=1}^{N}g_{i}\,\int d\{\bm{r}\}e^{\frac{i}{\hbar}(\ve{p}_{d}-\ve{p}'_{d})\cdot\bm{r}_{i}}\phi'(\{\bm{r}\})^{*}\phi(\{\bm{r}\})
\end{align}
\end{widetext}
where in the second line we introduced the compact notation $\{\bm{r}\}=\bm{r}_1,...\bm{r}_{N-1}$ and $\int d\{\bm{r}\}=\prod_{i}^{N-1}\int d\bm{r}_{i}$. In the third line we introduced the Fourier transform of the potential:
\begin{align}
\tilde{V}(\ve{q}) &= \int d\ve{x}\, e^{\frac{i}{\hbar}\ve{q}\cdot\ve{x}}\,V(\ve{x}),
\end{align}
and the function $\delta_{T}(x)$, which is defined as follows:
\begin{align}
\delta_{T}(x)=\frac{1}{2\pi\hbar}\int_{-T}^{T}d\tau \, e^{\frac{i}{\hbar} x\tau}\,  \qquad\quad& \lim_{T\rightarrow \infty} \delta_{T}(x) = \delta(x).
\end{align}
It is worth pointing out the different meaning of the two delta functions in Eq.~\eqref{eq:s1new}:
\begin{eqnarray}
\delta(\ve{p}_{d}+\ve{P}-\ve{p}'_{d}-\ve{P}') \;\;\;\;\;\;\;\textrm{and} \;\;\;\;\;\;\;\;\delta_{T}(E-E').\nonumber
\end{eqnarray}
The first one implies that the scattering conserves the total momentum, the second one is a Dirac delta only when $T\rightarrow \infty$ and in this limit it implies the conservation of energy.
Also, the following relations hold:
\begin{align}\label{delta2}
\lim_{T\rightarrow \infty} \delta_{T}(E)^{2} &=\delta(0)\delta(E) = \frac{T}{2\pi\hbar}\delta(E), \nonumber\\
\nonumber\\
\delta(\ve{p})^{2} &= \delta(0)\delta(\ve{p}) = \frac{\int_V d\ve{x}}{(2\pi \hbar)^{3}}\delta(\ve{p}),    
\end{align}
where the integral over the volume appearing in the second term of Eq.~\eqref{delta2} should be technically understood as over a very big but finite volume. \\
Inserting Eq.~\eqref{eq:s1new} in Eq.~\eqref{eq:Tprobability} and using the above relations, we obtain the scattering transition probability:
\begin{align}\label{eq:scattering probabilitynew}
&P_{in \rightarrow out}=\abs{\bra{\ve{p}_{d}',\ve{P}',\phi'}\hat{T}\ket{\ve{p}_{d},\ve{P},\phi}}^{2}&\nonumber\\
&= \frac{T\int_V d\ve{x}}{(2\pi \hbar)^8\hbar^{2}} \delta(E-E')\delta(\ve{p}_{d}+\ve{P}-\ve{p}'_{d}-\ve{P}')
 \left|\tilde{V}(\ve{p}_{d}-\ve{p}'_{d})\right|^{2} &\nonumber\\
 &\times \sum_{i,j=1}^{N}g_{i}g_{j}\int d\{\bm{r}\}\int d\{\bm{s}\} e^{\frac{i}{\hbar}(\ve{p}_{d}-\ve{p}'_{d})\cdot(\bm{r}_{i}-\bm{s}_{j})}& \nonumber\\
&\times  \;\phi'(\{\bm{r}\})^{*}\phi(\{\bm{r}\})\phi'(\{\bm{s}\})\phi(\{\bm{s}\})^{*}.& 
 \end{align}
From the last two lines of Eq.~\eqref{eq:scattering probabilitynew} we see that, in general, the contributions to the scattering probability given by each constituent of the target do not add coherently. However, there is a coherent effect when the contribution of each term is in phase with the others i.e. when we can approximate:
\begin{equation}\label{cohercond}
e^{\frac{i}{\hbar}(\ve{p}_{d}-\ve{p}'_{d})\cdot(\bm{r}_{i}-\bm{s}_{j})}\simeq1 
\end{equation}
in the fourth line of Eq.~\eqref{eq:scattering probabilitynew}.
Under this assumption and using the relation $\int d\{\bm{r}\}\phi'(\{\bm{r}\})^{*}\phi(\{\bm{r}\})=\delta_{\phi',\phi}$, Eq.~\eqref{eq:scattering probabilitynew} becomes:
\begin{eqnarray}\label{eq:fpio}
P_{in \rightarrow out} &=&
\frac{T\int_V d\ve{x}}{(2\pi \hbar)^8\hbar^{2}} \delta(E-E')\delta(\ve{p}_{d}+\ve{P}-\ve{p}'_{d}-\ve{P}') \nonumber\\
\nonumber\\
& \times&   \left|\tilde{V}(\ve{p}_{d}-\ve{p}'_{d})\right|^{2} \left(\sum_{i}^{N}g_{i}\right)^{2}\delta_{\phi',\phi}
\end{eqnarray}
Where $\delta_{\phi',\phi}$ implies that in this regime only scattering processes that do not excite the internal structure of the target are allowed.

Except for the delta and for the sum of the charges of the constituents there is no dependency on the internal structure of the target: the whole target behaves as a point-like object with total charge $G=\sum_{i=1}^{N}g_{i}$. \newline
The coherent scattering given by condition Eq.~\eqref{cohercond} is equivalent to requiring $|\ve{p}_{d}-\ve{p}'_{d}||\bm{r}_{i}-\bm{s}_{j}|\ll \hbar$. This means that on the one hand, the binding potential $U$ of the target must be strong enough to guarantee that the distances between its constituents are not too large; on the other hand the exchanged momentum $|\ve{p}_{d}-\ve{p}'_{d}|$ should not be too large.
However, $|\ve{p}_{d}-\ve{p}'_{d}|$ is constrained by the conservation of momentum and energy contained in the two deltas in Eq.~\eqref{eq:fpio}. This can be better seen by introducing the center of mass~\footnote{Note that $\ve{P}$ defined in Eq.~\eqref{eq:ptrr} is the momentum of the center of mass of the target alone, while now we have introduced $\ve{p}_{s}$, the momentum of the center of mass of the whole system, the target plus the incident particle.} ($\ve{p}_{s}$) and relative coordinates ($\ve{p}_{r}$) of the whole system:
\begin{align}\label{eq:prps}
\ve{p}_{r} = \frac{\ve{p}_{d}M-\ve{P}m_{d}}{m_{d}+M},& & \ve{p}_{s} = \ve{P}+\ve{p}_{d}
\end{align}
and by noticing that in the new coordinates, the two deltas can be written has:
\begin{align}\label{eq:dep}
&\delta(E-E')  = \delta\left(\frac{\ve{p}_{r}^{2}-{\ve{p}'}_{r}^{2}}{2m_{r}}+\frac{\ve{p}_{s}^{2}-{\ve{p}'}_{s}^{2}}{2m_{s}}\right), \nonumber\\ 
\nonumber\\
&\delta(\ve{p}_{d}+\ve{P}-\ve{p}'_{d}-\ve{P}') = \delta(\ve{p}_{s}-\ve{p}'_{s}),
\end{align}
where $m_{r}^{-1}= (M^{-1} +m_{d}^{-1})$ is the relative mass and $m_{s}= M+m_{d}$ is the total mass.
Using the restrictions given in Eq.~\eqref{eq:dep} together with the relations in Eq.~\eqref{eq:prps} we obtain:
\begin{align}
\abs{\ve{p}_{d}-\ve{p}'_{d}} = \abs{\ve{p}_{r}-\ve{p}'_{r}} &,& \abs{\ve{p}_{r}} = \abs{{\ve{p}'}_{r}},
\end{align}
which gives the following bound for $\abs{\ve{p}_{d}-\ve{p}_{d}'}$:
\begin{align}
\abs{\ve{p}_{d}-\ve{p}_{d}'} \le 2\abs{\ve{p}_{r}}.
\end{align} 
Given the above results, the coherent scattering regime in Eq.~\eqref{cohercond} is fulfilled when
\begin{equation}\label{qwe}
|\bm{r}_{i}-\bm{s}_{j}|\ll \frac{\hbar}{2|\ve{p}_{r}|},
\end{equation}
which means that the distance between the constituents of the target must be much smaller than the de Broglie wave length of the incident particle. This is in agreement with the result found for the simpler model discussed in section~\ref{sec-exact}.

\section{The Cross Section in the coherent scattering regime}\label{sec-cross}

We now provide an explicit expression for the cross section in the coherent scattering regime. In such a case, the transition probability $P_{in \rightarrow out}$ is given by Eq.~\eqref{eq:fpio}. Note that in the coherent scattering regime the scattering probability does not depend on the internal state $\phi(\{\bm{r}\})$. This is expected because, as already discussed, in this regime the incident particle does not resolve the internal structure of the target. 

The total cross section is defined as follow~\cite{Sakurai}:
\begin{align}\label{eq:cross section}
 &\sigma_{tot} = \nonumber\\
&=\frac{1}{n_{inc}}\sum_{\phi'}\int \frac{d P(\ve{p}'_{d},\ve{P}',\phi',\ve{p}_{d},\ve{P},\phi)}{dT} n(\ve{p}'_{d})n(\ve{P}')d\ve{p}'_{d} d\ve{P}'\nonumber\\
\end{align}
where $n(\ve{p}'_{d})=n(\ve{P}')=\frac{(2\pi\hbar)^{3}}{\int_V d\ve{x}}$ are the density of states, which we need to include because we are using plane waves in  Dirac normalization. The incident flux of particles $n_{inc}$ is:
\begin{align}
n_{inc}=\frac{1}{\int_V d\ve{x}} \frac{\abs{\ve{p}_{r}}}{m_{r}} 
\end{align}
where $\ve{p}_{r}$ is the relative momentum introduced in Eq.~\eqref{eq:prps} and $m_{r}$ the reduced mass defined just after Eq.~\eqref{eq:dep}.

Using Eq.~\eqref{eq:fpio} in Eq.~\eqref{eq:cross section}, we obtain for the total cross section:
\begin{align}\label{bo}
 \sigma_{tot}&=\frac{\left(\sum_{i=1}^{N}g_{i}\right)^{2}}{(2\pi)^{2}\hbar^{4}} \int{d\ve{p}'_{d}\,d\ve{P}'} \frac{m_{r}}{\abs{\ve{p}_{r}}} \delta(E-E')\times\nonumber\\
&\times\delta(\ve{p}_{d}-\ve{p}'_{d}+\ve{P}-\ve{P}')|\tilde{V}(\ve{P}'-\ve{P})|^{2}
\end{align}
It is convenient to introduce the center of mass $\ve{p}'_s$ and the relative $\ve{p}'_r$ final momenta:
\begin{equation}\label{prps}
 \mathbf{p}'_{r}=\frac{\mathbf{p}'_{d}M-\mathbf{P}'m_{d}}{m_{d}+M},\;\;\;\;\;\;\;\; \mathbf{p}'_{s}=\mathbf{P}'+\mathbf{p}'_{d}
\end{equation}
so that Eq.~\eqref{bo} becomes:
\begin{widetext}
\begin{eqnarray}\label{totalcrossnew}
\sigma_{tot}&=&\frac{m_{r}}{|\mathbf{p}_{r}|}\frac{\left(\sum_{i=1}^{N}g_{i}\right)^{2}}{(2\pi)^{2}\hbar^{4}}\int d\mathbf{p}'_{s}\, d\mathbf{p}'_{r}\delta(E_{s}+E_{r}-E'_{s}-E'_{r})
\delta(\mathbf{p}_{s}-\mathbf{p}'_{s})|\tilde{V}(\frac{M(\mathbf{p}'_{s}-\mathbf{p}{}_{s})}{\left(m_{d}+M\right)}+\mathbf{p}{}_{r}-\mathbf{p}'_{r})|^{2}\nonumber\\
\nonumber\\
&=&\frac{m_{r}}{|\mathbf{p}_{r}|}\frac{\left(\sum_{i=1}^{N}g_{i}\right)^{2}}{(2\pi)^{2}\hbar^{4}}\int d\mathbf{p}'_{r}\delta(E_{r}-E'_{r})|\tilde{V}(\mathbf{p}{}_{r}-\mathbf{p}'_{r})|^{2}\nonumber\\
\nonumber\\
&=&\frac{m_{r}^{2}}{\sqrt{E{}_{r}}}\frac{\left(\sum_{i=1}^{N}g_{i}\right)^{2}}{(2\pi)^{2}\hbar^{4}}\int d\Omega'_{r}\, dE'_{r}\sqrt{E'{}_{r}}\delta(E_{r}-E'_{r})|\tilde{V}(\sqrt{2m_{r}E{}_{r}}{\mathbf{n}}{}_{r}-\sqrt{2m_{r}E'{}_{r}}{\mathbf{n}}'{}_{r})|^{2}\nonumber\\
\nonumber\\
&=&m_{r}^{2}\frac{\left(\sum_{i=1}^{N}g_{i}\right)^{2}}{(2\pi)^{2}\hbar^{4}}\int d\Omega'_{r}|\tilde{V}(\sqrt{2m_{r}E{}_{r}}({\mathbf{n}}{}_{r}-{\mathbf{n}}'{}_{r}))|^{2}\nonumber\\
\end{eqnarray}
\end{widetext}
where we introduced the unitary vectors $\ve{n}_{r}:={\ve{p}_{r}}/{\abs{\ve{p}_{r}}}$ and $\ve{n}'_{r}:={\ve{p}'_{r}}/{\abs{\ve{p}'_{r}}}$. 

If the interaction potential depends only on the modulus of the relative distance, i.e. $V(\ve{x})=V(|\ve{x}|)$, then its Fourier transform depends only on the modulus of the transferred momentum $\tilde{V}(\ve{p}_{r}-\ve{p}'_{r})=\tilde{V}(|\ve{p}_{r}-\ve{p}'_{r}|)$. In such a case, we can write 
\begin{eqnarray}\label{nn}
\abs{\ve{n}_{r}-\ve{n}'_{r}}&=&\sqrt{{n}^{2}_{r}+{n}'^{2}_{r}-2\abs{\ve{n}_{r}}\abs{\ve{n}'_{r}}\cos\theta}\nonumber\\
\nonumber\\
&=&\sqrt{2(1-\cos\theta)}=2\sin\left(\theta/2\right)
\end{eqnarray}
where $\theta$ denotes the angle between $\ve{n}_{r}$ and $\ve{n}'_{r}$. Choosing a reference system where $\ve{n}_{r}$ is oriented along the $z$-axis, we can rewrite $d\Omega_{r}' = d\cos\theta d\varphi$ and then obtain:
\begin{eqnarray}\label{totalcrossnew2}
\sigma_{tot}\!\!&=&m_{r}^{2}\frac{\left(\sum_{i=1}^{N}g_{i}\right)^{2}}{(2\pi)^{2}\hbar^{4}}\int_{0}^{2\pi} d\varphi\int_{0}^{\pi} d\theta\sin(\theta)\times\\
&&\nonumber\\
\!\!&\times&|\tilde{V}(\sqrt{8m_{r}E{}_{r}}\sin\left(\theta/2\right))|^{2}\nonumber\\
&&\nonumber\\
\!\!&=&m_{r}^{2}\frac{\left(\sum_{i=1}^{N}g_{i}\right)^{2}}{2\pi\hbar^{4}}\!\!\int_{-1}^{1}\!\! d(\cos\theta)|\tilde{V}(\sqrt{8m_{r}E{}_{r}}\sin\left(\theta/2\right))|^{2},\nonumber
\end{eqnarray}
from which we get the following expression for the differential cross section:
\begin{equation}\label{diffcrossnew}
\frac{d \sigma_{tot}}{d(\cos\theta)}=\left(\sum_{i=1}^{N}g_{i}\right)^{2}\frac{m_{r}^{2}}{2\pi\hbar^{4}}|\tilde{V}(\sqrt{8m_{r}E{}_{r}}\sin\left(\theta/2\right))|^{2}.
\end{equation}
As already discussed in the previous sections, the cross section in Eq.~\eqref{diffcrossnew} depends on the total charge of the target $G=\sum_{i=1}^{N}g_{i}$. Therefore, the cross section can be easily increased quadratically.


\section{Rutherford Scattering}\label{sec-ruth}

In Rutherford's experiment, $\alpha$ particles were fired to a thin foil of different materials and the angular distribution of the scattered particle was measured. The experiments showed clearly that there is a non negligible probability for the $\alpha$ particles of being scattered at big angles. These observations leaded Rutherford to suggest the planetary model of the atom. The connection between Rutherford scattering and coherent scattering is that, in order to fit the experimental data with his model, Ruthrford treated both the $\alpha$ particles and the nucleus of atoms as point like object, without considering the details of their internal structure. Therefore, in Rutherford scattering, all the protons of a nucleus scatter coherently the incident $\alpha$ particle. 

We new derive the Rutherford cross section starting from the coherent cross section of Eq.~\eqref{diffcrossnew}. The interaction is described by the Coulomb potential $V(\ve{x})=1/|\ve{x}|$ and its Fourier transform is:
\begin{align}\label{eq:fcp}
\tilde{V}(\ve{p}'_{r}-\ve{p}_{r}) &= \int d\ve{x} \frac{e^{\frac{i}{\hbar}(\ve{p}'_{r}-\ve{p}_{r})\cdot \ve{x}}}{\abs{\ve{x}}} = \frac{4\pi \hbar^{2}}{(\ve{p}_{r}-\ve{p}'_{r})^{2}}\\
&=\frac{4\pi \hbar^{2}}{\abs{\ve{p}_{r}}^{2}+\abs{\ve{p}'_{r}}^{2} -2\abs{\ve{p}'_{r}}\abs{\ve{p}_{r}}\cos\theta}\nonumber\\ 
&= \frac{\pi \hbar^{2}}{m_{r}E_{r}}\frac{1}{(1-\cos\theta)}=\frac{ \pi \hbar^{2}}{2m_{r}E_{r}}\frac{1}{\sin^{2}(\theta/2)}\nonumber
\end{align}
where, in the third line, we used $|\ve{p}_{r}|=|\ve{p}'_{r}|$. With the help of Eq.~\eqref{eq:fcp} we can rewrite Eq.~\eqref{diffcrossnew} as follows:
 \begin{align}
 \frac{d \sigma_{tot}}{d(\cos\theta)}=\left(\sum_{i=1}^{Z}g_{i}\right)^{2}\frac{\pi}{8 E_{r}^{2} \sin^{4}(\theta/2)}
 \end{align}
 where $g_{i}$ is the Coulomb coupling constant between the scattered $\alpha$ particle and the $i$-th proton of the nucleus:
 \begin{align}\label{eq:charge}
 g_{i}= \frac{q_{d}q_{i}}{4\pi\epsilon_{0}}=\frac{e^2}{2\pi\epsilon_{0}}.
 \end{align}
In Eq.~\eqref{eq:charge} $\epsilon_{0}$ is the vacuum permittivity $q_{d}=2e$ is the charge of the $\alpha$ particles and $q_{i}=e$ are the charges of the protons of the nucleus. Since $i$ labels the different protons of the nucleus, then $i=1,2,...Z$.
Therefore we obtain:
\begin{align}\label{eq:rutherford}
\frac{d \sigma_{tot}}{d(\cos\theta)}=\frac{Z^2e^{2}}{32 \pi \epsilon_{0}^{2}E_{r}^{2}\sin^{4}(\theta/2)}.
\end{align}
Since the mass of the $\alpha$ particle is much smaller then that of the nucleus, we can approximate $E_{r} \simeq E_{\alpha}$ and $\theta\simeq \theta_{\alpha}$ where $E_{\alpha}$ and $\theta_{\alpha}$ denote, respectively, the energy and the scattering angle of the incident $\alpha$ particle. Then Eq.~\eqref{eq:rutherford} becomes:
\begin{align}\label{eq:rutherford2}
\frac{d \sigma_{tot}}{d(\cos\theta_{\alpha})}=\frac{Z^2e^{2}}{32 \pi \epsilon_{0}^{2}E_{\alpha}^{2}\sin^{4}(\theta_{\alpha}/2)},
\end{align}

which is the Rutherford cross section. Because of the coherence of the scattering process, the cross section is proportional to the square of the number of protons of each nucleus~\footnote{In the original paper of Rutherford~\cite{rutherford2012scattering} it was supposed that the total cross section was proportional to $A^2$ instead of $Z^2$. This is due to the fact that at the time when Rutherford performed the experiment, the existence of neutrons was not known. However, since now we know that only protons interact electromagnetically with the $\alpha$ particles, the cross section is proportional only to the square of the number of protons $Z$}. This dependence has been observed experimentally by Rutherford and his collaborators~\cite{Geiger, rutherford2012scattering}. In their experiments, they measured the number of scintillations per minute in their detector, which is proportional to the number of $\alpha$ particles scattered per minute. Therefore, according to Eq.~\eqref{eq:rutherford2}, when the experiment is repeated with different materials one should expect the quantity $N/Z^2$ to be constant. However, there is an additional effect, which must be take into account: when an $\alpha$ particle goes through the material it can also be absorbed. In such a case the $\alpha$ particle is not detected. Bragg and his collaborators showed that the thickness of the layer an alpha particle can go through without being absorbed, goes as the inverse of the square root of the mass number $A$. Therefore, when making the comparison with experimental data, also this effect must be included. In such a case the quantity $N\sqrt{A}/Z^2$ is expected to be constant. The results of this analysis is reported in Table I. From the last column of the table we see that, apart for aluminum, for all the others materials we obtain values which are similar, as expected. 
\begin{table}[!]
\begin{tabular}{|l| c|c|c|c|}
\hline
Material	& $A$ 		& $Z$ &   $N$ & 	$N\sqrt{A}/{Z^2}$  \\\hline\hline
Lead 		& $207$ 	& $82$ & $62$ & 	$0,13$  \\\hline
Gold 		& $197$ 	& $79$ & $67$ & 	$0,15$  \\\hline
Platinum 	& $195$ 	& $78$ & $63$ & 	$0,14$  \\\hline
Tin 		& $119$ 	& $50$ & $34$ & 	$0,15$  \\\hline
Silver 		& $108$ 	& $47$ & $27$ & 	$0,13$  \\\hline
Copper 	& $64$ 	& $29$ & $14,5$ &	$0,14$  \\\hline
Iron 		& $56$ 	& $26$ & $10,5$ & 	$0,12$  \\\hline
Aluminum 	& $27$ 	& $13$ & $3,4$ & 	$0,10$  \\\hline
\end{tabular}
\caption{In the above table we show, for different materials, the mass number $A$ (i.e. the total number of protons and neutrons of each atom), the atomic number $Z$ and the number of scintillations observed per minute $N$ as reported in~\cite{rutherford2012scattering}, which is proportional to the cross section. In the last column we compute $N\sqrt{A}/{Z^2}$, which is expected to be constant.}
\end{table}

\section{Conclusions}\label{sec-conc} 

In this paper we have explicitly shown that a sufficient condition for coherent scattering is:
\begin{align}\label{eq:lambda_cond}
\lambda_{r}\gg  L,
\end{align}
with $\lambda_{r}$ the de Broglie wave-length associated to the incident particle, as seen in the reference frame of the target, and $L$ the typical spatial extension of the target.
In this regime, the differential cross section in Born approximation is given by Eq.~\eqref{diffcrossnew}:
\begin{equation}
\frac{d \sigma_{tot}}{d(\cos\theta)}=\left(\sum_{i=1}^{N}g_{i}\right)^{2}\frac{m_{r}^{2}}{2\pi\hbar^{4}}|\tilde{V}(\sqrt{8m_{r}E{}_{r}}\sin\left(\theta/2\right))|^{2},
\end{equation}
where the coherent effects are embedded in the dependence of the cross section on the square of the total charge $G=\sum_{i}^{N}g_{i}$.

\section*{Acknowledgements}
The authors acknowledge financial support from the EU project NANOQUESTFIT, INFN, the John Templeton foundation (grant 39530), the University of Trieste (grant FRA 2013) and the COST Action MP1006. The authors thank Prof. M. Arndt, Prof. K. Hornberger, Prof. N. Paver, Prof. R. Rui, Prof. D. Treleani and Prof. E. Ulbricht for very enjoyable and stimulating discussions on this topic.

\section*{Appendix A: One Dimensional 2-Delta potential Scattering } 
In this appendix we show explicitly how to derive the reflectivity $R$ of Eq.~\eqref{riflesso}.
We first solve the equation for the stationary states: 
\begin{equation}\label{eigenprob}
-\frac{\hbar^{2}}{2m}\frac{d^{2}}{dx^{2}}\psi\left(x\right)+\alpha\left[\delta\left(x\right)+\delta\left(x-a\right)\right]\psi\left(x\right)=E\psi\left(x\right)
\end{equation}
associated to the Hamiltonian in Eq.~\eqref{eq:HHH}, where $m$ is the mass of the particle and $\alpha$ is the coupling constant of the $\delta$-potential (with dimensions
$\left[\alpha\right]=$J m). 
To solve Eq.~\eqref{eigenprob}, we divide the $x$-axis in three regions: region ``1'' with $x\in(-\infty,0)$, which is the region on the left of the first delta, region ``2'' with $x\in (0, a)$, which is the region between the two deltas and region ``3'' with $x \in (a, \infty)$, which is the region on the right of both deltas. We compute the reflection probability of a wave packet, coming from the left, which scatters off the delta potentials. Following the same procedure used in~\cite{cald}~\footnote{In \cite{cald} the calculation with a wave packets is performed for a square well potential. However the reasoning there is valid also for the problem we are considering.}, it can been shown that the reflection probability can be computed by working with plane waves instead of considering wave packets. For each region we introduce the corresponding plane wave
solutions of the free Schr{\"o}dinger equation, $\psi_{j}\left(x\right)$ with
$j=1,2,3$, which have the form:
\begin{equation}
\psi_{j}\left(x\right)=c_{j}e^{ikx}+d_{j}e^{-ikx}\;\;\;\;\textrm{with}\;\;\;\; k=\frac{\sqrt{2mE}}{\hbar}.
\end{equation}
The reflection probability is given by 
\begin{equation}\label{Rapp}
R:=|d_1|^2/|c_1|^2,
\end{equation}
therefore we need to determine only the coefficients $c_{j}$ and $d_{j}$. This is done by using the following constrains imposed on the wave function $\psi(x)$ by the Dirac-delta potentials:
\begin{enumerate}

\item The wave function must be continuous at the origin and at the point $a$, i.e.: 
\begin{equation}
\psi_{1}\left(0\right)=\psi_{2}\left(0\right)\;\;\;\;\textrm{and}\;\;\;\;\psi_{2}\left(a\right)=\psi_{3}\left(a\right)\label{cond1}
\end{equation}

\item Given a Dirac delta potential at a point $a$, the right and left derivatives of the wave function at that point must obey the constraint:
\begin{equation}
\psi'_{R}\left(a\right)-\psi'_{L}\left(a\right)=\frac{2m\alpha}{\hbar^{2}}\psi_{L}\left(a\right)
\end{equation}
where $\psi'_{R/L}\left(x\right):=\frac{d}{dx}\psi_{R/L}\left(x\right)$.
In our case this corresponds to the following condition at the origin:
\begin{equation}
\psi'_{2}\left(0\right)-\psi'_{1}\left(0\right)=\frac{2m\alpha}{\hbar^{2}}\psi_{1}\left(0\right)\label{cond2}
\end{equation}
and the one around $a$: 
\begin{equation}
\psi'_{3}\left(a\right)-\psi'_{2}\left(a\right)=\frac{2m\alpha}{\hbar^{2}}\psi_{2}\left(a\right).\label{cond3}
\end{equation}

\end{enumerate}
Eqs.~\eqref{cond1}, \eqref{cond2} and \eqref{cond3} set four conditions on the wave function $\psi(x)$. Since, $\psi(x)$ depends on the six parameters $c_j$ and $d_j$ ($j=1,2,3$), we still have two degrees of freedom. However, because we are considering particles coming from the left, and $|c_1|^2$ and $|d_3|^2$ give, respectively, the flux of incoming particles from the left and from the right, we can set $c_{1}=1$ and $d_{3}=0$. 

Therefore we have:
\begin{eqnarray}
\psi_{1}\left(x\right)&=&e^{ikx}+d_{1}e^{-ikx}\, , \nonumber\\
\nonumber\\
\psi_{2}\left(x\right)&=&c_{2}e^{ikx}+d_{2}e^{-ikx}\, ,\nonumber\\
\nonumber\\
\psi_{3}\left(x\right)&=&c_{3}e^{ikx}\, .\nonumber
\end{eqnarray}
Using the conditions given in Eqs.~(\ref{cond1}), (\ref{cond2}) and (\ref{cond3}) we get: 
\begin{eqnarray}
d_{2}&=&-\frac{i\beta e^{2ika}}{e^{2ika}+2i\beta+\beta^{2}-1}\,,\\
\nonumber\\
c_{2}&=&\frac{\beta(i+\beta)}{e^{2ika}+2i\beta+\beta^{2}-1}\, , \\
d_{1}&=&-\frac{-1+i\beta+e^{2ika}(1+i\beta)}{e^{2ika}+(i+\beta)^{2}} \, , \\
\nonumber\\
c_{3}&=&c_{2}+d_{2}e^{-2ika} \, .
\end{eqnarray}
where we introduced $\beta=\frac{k\hbar^{2}}{m\alpha}$.
The reflection probability $R$ defined in Eq.~(\ref{Rapp}) than correspond ti Eq.~\eqref{riflesso}
\section*{Appendix B: Wave packets instead of plane waves}

We extend the conditions for coherent scattering  derived in section~\ref{sct} by using plane waves, to the case where the initial state of the incident particle is a wave packet. 

For the sake of simplicity, let us work in the reference frame where the target is at rest. Then the inequality in Eq.~\eqref{qwe}, which guarantees coherent scattering for plane waves, becomes: 
\begin{align}\label{eq:1beamcoherent}
\abs{\ve{p}_{d}} \ll L^{-1} \hbar,
\end{align}
where $L$ is the spatial extension of the target and $\ve{p}_{d}$ is the momentum of the incident particle, before the scattering.

Now, instead of a plane wave with definite momentum, suppose we have a wave packet $\ket{\psi}$. We expand it in the plane waves basis:
\begin{align}
\ket{\psi} = \int d\ve{p} \ket{\ve{p}}{\psi}(\ve{p})
\end{align}
In order to have coherent scattering, all plane waves composing the wave packet $\ket{\psi}$ need to fulfill inequality \eqref{eq:1beamcoherent}. This is guaranteed when:
\begin{align}\label{eq:p_ineq}
 \abs{\ve{p}_{max}}  \ll L^{-1}\hbar
\end{align}
where $\ve{p}_{max}$ denotes the maximum significant momentum of the wave packet. For a wave packet with momentum average $\langle \hat{\ve{p}}\rangle_{\psi}$ and  momentum spread $\Delta \ve{p}_{\psi}$, if we disregard the contributions coming from the tails of  the momentum distribution $\psi(\ve{p})$, we can approximate $\abs{\ve{p}_{max}} \simeq \abs{\langle \hat{\ve{p}} \rangle_{\psi}} + \Delta \ve{p}_{\psi}$. Then the condition for coherent scattering becomes:
\begin{align}\label{eq:coherent_con}
\abs{\langle \hat{\ve{p}} \rangle_{\psi}} + \Delta \ve{p}_{\psi} \ll L^{-1}\hbar
\end{align}  
or equivalently:
\begin{align}
\Delta \ve{x}_{\psi} \gg L
\end{align}  
as one can easily prove with the help of the uncertainty principle.

\section*{Appendix C: Ensembles of wave packets}
Now we consider the case of a incident beam of particles described by different wave packets.
This beam is then described by a statistical mixture:  
\begin{align}
\hat{\rho} = \sum_{i} \lambda_{i} \ket{\psi_{i}}\bra{\psi_{i}}
\end{align}
with $\lambda_i$ are weights which sum to one, and $\ket{\psi_{i}}$ are wave packets, each describing a single particle state. If all wave packets $\ket{\psi_{i}}$ fulfill the coherent scattering condition \eqref{eq:coherent_con}, then the beam described by $\hat{\rho}$ also describes coherent scattering, and verifies the following inequality:
\begin{align}\label{eq:coherent_ineq}
\langle\hat{\ve{p}}\rangle_{\hat{\rho}}^{2} + (\Delta \ve{p}_{\hat{\rho}})^{2} \ll L^{-2} \hbar^{2}
\end{align}
in fact:
\begin{eqnarray}\label{eq:coherent_ineq2}
\Delta \ve{p}_{\hat{\rho}}^{2}+\langle\hat{\ve{p}}\rangle_{\hat{\rho}}^{2} &=&\textrm{Tr}\left[\mathbf{\hat{p}}^{2}\hat{\rho}\right] =\sum_{i}\lambda_{i}\left\langle \hat{p}^{2}\right\rangle _{\psi_{i}}\nonumber\\
\nonumber\\
&=&\sum_{i}\lambda_{i}\left[(\Delta \ve{p}_{\psi_{i}} )^{2}+\left\langle \hat{\ve{p}}\right\rangle _{\psi_{i}}^{2}\right]\nonumber\\
\nonumber\\ 
&\ll& \hbar^2 L^{-2} \sum_{i}\lambda_{i} = \hbar^2 L^{-2} 
\end{eqnarray}
Now the following apparent paradox appears. In general, an ensemble of ``large enough" wave packets (large with respect to $L$), all of which scatter coherently, can be equivalently described by an ensemble of ``small" wave packets. Equivalent means that the two ensembles are associated to the same density matrix $\hat{\rho}$. In particular, the second ensemble can be chosen in such a way that the size of the wave packets is small to the point that each incident particle sees only one constituent of the target. In such a case there cannot be coherent scattering. Here is the paradox: the two ensembles predict a different behavior for the scattering process, and this cannot be true, because they are equivalent and according to quantum theory they must give the same result. We clarify the situation. More specifically, we show that any $\hat{\rho}$ which satisfies the coherent scattering condition in Eq.~\eqref{eq:coherent_ineq} cannot be decomposed in terms of  wave packets with spread in position smaller than $L$. 

In fact, suppose that:
\begin{align}
\hat{\rho} = \sum_{i} w_{i} \ket{\varphi_{i}}\bra{\varphi_{i}}
\end{align}
such that every $\ket{\varphi_{i}}$ satisfies: $\Delta \ve{x}_{\varphi_{i}}  < L $. Then because of the uncertainty principle:
\begin{align}\label{eq:ineq_cond}
\Delta \ve{p}_{\varphi_{i}} > \hbar L^{-1}.
\end{align}
Of course this is in contrast with the coherent scattering condition \eqref{eq:coherent_con}.
The important point is that, in this case, the statistical mixture $\hat{\rho}$ verifies the following inequality:
\begin{align}\label{eq:cond2}
\left\langle \hat{\ve{p}}\right\rangle _{\hat{\rho}}^{2}+(\Delta \ve{p}_{\rho})^{2} > h^{2}L^{-2}
\end{align}
as a simple computation similar to the one which brings from Eq.~\eqref{eq:coherent_ineq} to Eq.~\eqref{eq:coherent_ineq2} shows.
As we can see, conditions \eqref{eq:coherent_ineq} and  \eqref{eq:cond2} are not compatible, meaning that the request of coherent scattering for $\hat{\rho}$ is incompatible with the request that the statistical operator  can be decomposed in a mixture of wave functions $\ket{\varphi_{i}}$, each of which is small enough to see only one scattering center.



\end{document}